\documentclass[preprint,preprintnumbers, prd, floatfix,  superscriptaddress,nofootinbib] {revtex4}
\usepackage{graphicx}
\usepackage{epsfig}
\usepackage{bm}
\usepackage{amssymb}
\usepackage{float}
\usepackage{amsmath}
\usepackage{dcolumn}
\usepackage{cancel}
\usepackage[colorlinks]{hyperref}
\usepackage[usenames,dvipsnames]{color}
\usepackage{epstopdf}
\hypersetup{
     breaklinks=true,
    pdfstartview={FitH},    colorlinks=true,       
    linkcolor=blue,          
    citecolor=red,        
    filecolor=magenta,      
    urlcolor=blue,           
    anchorcolor=green,      
    linktocpage=true
}

\def\doi{http://doi.org}

\begin{document}

 \title{Cosmological perturbation and matter power spectrum in bimetric massive gravity}

 \author{Chao-Qiang Geng}
\email{geng@phys.nthu.edu.tw}
\affiliation{
Synergetic Innovation Center for Quantum Effects and Applications (SICQEA),
Hunan Normal University, Changsha 410081, China}
 \affiliation{National Center for Theoretical Sciences, Hsinchu,
Taiwan 300}
\affiliation{Department of Physics, National Tsing Hua University,
Hsinchu, Taiwan 300}

\author{Chung-Chi Lee}
\email{lee.chungchi16@gmail.com}
\affiliation{
DAMTP, Centre for Mathematical Sciences, University of Cambridge, Wilberforce Road, Cambridge CB3 0WA, UK}
 \affiliation{National Center for Theoretical Sciences, Hsinchu,
Taiwan 300}

\author{Kaituo Zhang}
\email{ktzhang@ahnu.edu.cn}
\affiliation{Department of Physics, Anhui Normal University, Wuhu, Anhui 241000, China}
 \affiliation{National Center for Theoretical Sciences, Hsinchu,
Taiwan 300}

\begin{abstract}

We discuss the linear perturbation equations with the synchronous gauge in a minimal scenario of the bimetric massive gravity theory.
We find that the matter density perturbation  and matter power spectrum are suppressed. We also examine the ghost and stability problems and show that the allowed deviation of this gravitational theory from the cosmological constant is constrained to be smaller than $\mathcal{O}(10^{-2})$ by the large scale structure observational data.

\end{abstract}

\maketitle

\section{Introduction} \label{sec:introduction}

As the cosmological observations~\cite{Perlmutter1999, Riess1998,Tegmark2004,Eisenstein2005,Spergel} have shown that our universe is undergoing an accelerated cosmic expansion, the Einstein equation in general relativity (GR) without the mystery cosmological constant term needs to be modified.
Among these modified gravity theories, the DGP braneworld~\cite{Dvali2000} and massive gravity theories can both explain the late-time acceleration without adding dark energy additionally.

Massive gravity with a massive spin-2 field was first constructed by Fierz and Pauli (FP)~\cite{Fierz1939} in 1939.
However, van Dam, Veltman and  Zakharov(vDVZ) found that
in the zero limit of the graviton mass, the modification of the Newtonian potential is not continuous ~\cite{Dam1970,Zakharov1970,Iwasaki1970}, resulting in a large correction to the bending of light around the sun, which mismatches the current solar system observations.
To resolve this problem, Vainshtein~\cite{Vainshtein1972} proposed a mechanism of nonlinear interactions, which could recover GR in the zero graviton mass limit, which is applicable to many modified gravitational theory, such as the DGP braneworld and Galileon theories~\cite{Lue:2004rj, Koyama:2007ih, Koyama:2013paa, Barreira:2014zza, Koyama:2015oma, Koyama:2015vza, Crisostomi:2017lbg, Koyama:2009me}.
Nevertheless, the massive gravity also
suffers from the Boulware Deser (BD) ghost instability at the same time~\cite{Boulware1972}.

Recently, de Rham, Gabadadze and Tolley (dRGT)~\cite{Rham2010,Rham2011} have successfully built the covariant theory of massive gravity by introducing a second reference metric in addition to the ordinary one, which can be fully BD ghost-free to both linear and non-linear orders~\cite{Hassan20123}. Nevertheless, the homogeneous and isotropic cosmological  solutions are not stable in such a non-linear theory~\cite{Amico2011, Felice2012,  Rham2014}.
To solve this problem, Hassan and Rosen have extended massive gravity with a non-dynamic second metric to bigravity with a dynamic one~\cite{Hassan20121,Hassan20122}.
Besides being BD ghost-free~\cite{Hassan20123} for massive gravitons, there exist some solutions in this bimetric theory with interacting massless and massive spin-2 fields.
Furthermore, this model can be applicable to realize the late-time accelerated cosmic expansion without a cosmological constant~\cite{Akrami2013,Volkov2013}.
As the theory contains six free coupling parameters, there are various possible cosmological solutions with different choices of these couplings.
However, many bigravity models encounter with Higuchi ghosts~\cite{Higuchi:1986py,Higuchi:1989gz} or gradient instabilities under cosmological perturbations~\cite{Crisostomi2012, Solomon2014, Koenning2014, Lagos2014, Comelli2014prd, Koenning2014prd2, Berg2012}.
There are several attempts to find viable cosmological solutions~\cite{Lagos2016JCAP, Koenning20141, Akrami2014, Felice2014, Akrami20131, Rham2015, Strauss2012, Volkov2012,Katsuragawa2014}.
In Ref.~\cite{Crisostomi2012}, it is shown that the instabilities can be avoided by introducing a mirror dark matter sector minimally coupled to the second metric. In Ref.~\cite{Akrami2014}, the authors find that, in the case without the cosmological constant related parameter $\beta_0$,
the stability condition requires that
the Planck mass  of the second metric
is much less than that of the first one.
In this study, we examine whether the absence of instabilities  can occur in a  minimal case with the parameter $\beta_0$ but without the mirror dark matter term.
In particular, we use the general conditions  given in Ref.~\cite{Koenning2015} to avoid the Higuchi ghosts and gradient instabilities at the linear level.
Some recent studies on bigravity can be seen in Refs.~\cite{Mortsell2017, Lagos2016, Apolo2016, Sakakihara2016, Kimura2016, Do2016,Aoki2016, Gao2016, Darabi2016, Akrami2015, Hassan2014, Rham20142, Cusin2014, Noller2015, Comelli2014, Rham20143} and references therein.

As demonstrated in the literature\footnote{See Ref.~\cite{Ade:2015rim}, for example.},
the modified gravity theories without the $\Lambda$CDM approach are hard to fit the cosmic microwave background radiation (CMB) and  large scale structure observational data.
In order to check the viability of the bigravity model,
it is helpful to explore the feature of the linear density perturbations as well as the matter power spectrum in a viable non-trivial minimal scenario, which has the $\Lambda$CDM limit associated with a modification term.
This type of the study  has been done in the DGP~\cite{Falck:2014jwa} and Galileon frameworks~\cite{Barreira:2013eea, Li:2013tda, Barreira:2013xea}.
In this work,
%
%
we plan to estimate
the allowed window of the model parameter by comparing with our numerical calculations to the most
recent observational data, and  examine the stability with the constrained parameter.
Our goal is to find out whether there are still some possible deviations between our viable minimal scenario of the
bimetric massive gravity theory and the $\Lambda$CDM model.

This paper is organized as follows.
In Sec.~\ref{sec:model} we derive the effective equations of dark energy at the  background level.
In Sec.~\ref{sec:density-pert}, we consider scalar perturbations in the synchronous gauge and obtain the perturbed equations.
We numerically solve the linear perturbation equations in Sec.~\ref{sec:mpk}.
We discuss the ghost and stability problems and show the matter power spectrum in Sec.~\ref{sec:stability}.
Finally, the conclusion is given in Sec.~\ref{sec:conclusion}.

\section{Background of bimetric massive gravity}
\label{sec:model}

In this section, we would first review some basic background of the bimetric massive gravity theory and then
introduce our minimal scenario for the theory.
We start with
the action of the theory, given by~\cite{Hassan20121},
\begin{eqnarray}
\label{eq:action}
\mathcal{S} =- \int d^4 x \left( \frac{\sqrt{-g}}{16 \pi G} R(g) + \frac{\sqrt{-f}}{16 \pi G_f} R(f) \right)+  m^2\int d^4 x \frac{\sqrt{-g}}{8 \pi G} \sum\limits_{n=0}^4 \beta_n e_n \left(\mathbb{X} \right) + \mathcal{S}_M(g_{\mu\nu}, \Psi) \,,
\end{eqnarray}
where $R(g)$ and $R(f)$ are the Ricci scalars, corresponding to the ordinary and new metrics $g_{\mu \nu}$ and $f_{\mu \nu}$, respectively, $G_f$ is the gravitational constants for the new metric, $m$ is a mass parameter, $\mathcal{S}_M(g_{\mu\nu}, \Psi)$ is the action of the matter term with the matter field $\Psi$, $\beta_n$ are the arbitrary constants, and $e_n$ are defined by,
\begin{eqnarray}
\label{eq:polynomial}
&& e_0\left(\mathbb{X}\right) = 1 \,, \qquad e_1\left(\mathbb{X}\right) = \left[\mathbb{X}\right] \,,  \qquad e_2\left(\mathbb{X}\right) = \frac{1}{2} \left( \left[\mathbb{X}\right]^2 - \left[\mathbb{X}^2\right] \right) \,, \nonumber \\
&& e_3\left(\mathbb{X}\right) = \frac{1}{6} \left( \left[\mathbb{X}\right]^3 -3 \left[\mathbb{X}\right] \left[\mathbb{X}^2\right] + 2 \left[\mathbb{X}^3\right] \right) \,, \qquad e_4\left(\mathbb{X}\right) = \det \mathbb{X} \,,
\end{eqnarray}
with $\mathbb{X}= \sqrt{g^{\alpha \beta} f_{\beta \gamma}}$.
For convenience, we will absorb $m^2$ into $\beta_n$, set $8\pi G_f=1$~\cite{Khosravi:2012rk, Konnig:2014xva}, and denote the trace of the matrix $\mathbb{X}$ by  $\left[ \mathbb{X} \right]$.

Varying the action in Eq.~(\ref{eq:action}) with respect to $g_{\mu \nu}$ and $f_{\mu \nu}$, the modified Einstein equations can be derived to be
\begin{eqnarray}
\label{eq:fieldeq1}
&& G_{\mu \nu} + \sum\limits_{n=0}^3 (-1)^n\beta_n g_{\mu \lambda} (J_n)^\lambda_\nu = \kappa^2 T^M_{\mu\nu} \,, \\
\label{eq:fieldeq2}
&& F_{\mu \nu} +\sum\limits_{n=0}^3 (-1)^n\beta_{4-n} f_{\mu \lambda} (J_n)^\lambda_\nu = 0 \,,
\end{eqnarray}
respectively, where $\kappa^2 = 8 \pi G =1$, $G_{\mu \nu}$ ($F_{\mu \nu}$) is the Einstein tensor for the metric $g_{\mu \nu}$ ($f_{\mu \nu}$), $T^M_{\mu\nu}$ is the energy-momentum tensor, and $(J_n)^{\lambda}_\nu$ are defined by
\begin{eqnarray}
\label{eq:polynomial2}
&& J_0 = \mathbb{I} \,, \qquad J_1 = \mathbb{X} -\mathbb{I} \left[\mathbb{X}\right] \,, \qquad J_2 = \mathbb{X}^2 - \mathbb{X} \left[\mathbb{X}\right] + \frac{1}{2} \mathbb{I}\left( \left[\mathbb{X}\right]^2 - \left[\mathbb{X}^2\right] \right) \,, \nonumber \\
&& J_3 = \mathbb{X}^3 - \mathbb{X}^2 \left[\mathbb{X}\right] + \frac{1}{2} \mathbb{X} \left( \left[\mathbb{X}\right]^2 - \left[\mathbb{X}^2\right] \right) - \frac{1}{6}\mathbb{I} \left( \left[\mathbb{X}\right]^3 - 3 \left[\mathbb{X}\right] \left[\mathbb{X}^2\right] + 2 \left[\mathbb{X}^3\right] \right) \,,
\end{eqnarray}
with $\mathbb{I}$  the identity matrix.

By taking the Friedmann-Lema\"itre-Robertson-Walker (FLRW) types of the metric~\cite{Konnig:2014xva},
\begin{eqnarray}
\label{eq:g_bg}
&& ds^2 = g_{\mu \nu} dx^{\mu} dx^{\nu} = -dt^2 + a(t)^2 dx^i dx_i \\
\label{eq:f_bg}
&& ds_f^2 = f_{\mu \nu} dx^{\mu} dx^{\nu} = -\frac{\dot{b}^2}{\dot{a}^2} dt^2 + b(t)^2 dx^i dx_i \\
\end{eqnarray}
the Friedmann equations are given by
\begin{eqnarray}
\label{eq:eom1}
&& H^2 = \frac{1}{3} \left( \rho_M + \beta_0 + 3 \beta_1 \frac{b}{a} + 3 \beta_2 \frac{b^2}{a^2} + \beta_3 \frac{b^3}{a^3} \right) \,, \\
\label{eq:eom2}
&& \dot{H} = -\frac{1}{2} \left( \rho_M + P_M +  \beta_1 \frac{b}{a}+2\beta_2\frac{b^2}{a^2}+\beta_3{b^3}{a^3}-\beta_1\frac{\dot{b}}{\dot{a}}-2\beta_2\frac{b}{a}\frac{\dot{b}}{\dot{a}}-\beta_3\frac{b^2}{a^2}\frac{\dot{b}}{\dot{a}} \right) \,,
\end{eqnarray}
and
\begin{eqnarray}
\label{eq:eom3}
&& H^2 = 
\frac{1}{3}\frac{a}{b} \left( \beta_1  + 3 \beta_2\frac{b}{a} + 3 \beta_3 \frac{b^2}{a^2} + \beta_4 \frac{b^3}{a^3} \right) \,, \\
\label{eq:eom4}
&& H^2+2\frac{H}{H_f}\frac{\ddot{a}}{a}= \left(\beta_2+2\beta_3\frac{b}{a}+\beta_4\frac{b^2}{a^2}+\beta_1 \frac{\dot{a}}{\dot{b}}+2\beta_2\frac{b}{a}\frac{\dot{a}}{\dot{b}}+\beta_3\frac{b^2}{a^2}\frac{\dot{a}}{\dot{b}} \right)\,\,,
\end{eqnarray}
for $g_{\mu\nu}$ and $f_{\mu\nu}$, respectively, where  $H_{(f)}=\dot{a}/a \ (\dot{b}/b)$ is the Hubble constant of $g_{\mu\nu}$ ($f_{\mu\nu}$), and $\rho_M = \rho_r + \rho_m$ ($P_M = P_r + P_m$) is the energy density (pressure) of the radiation $\rho_r$ ($P_r$) and matter $\rho_m$ ($P_m$).

\begin{table}[htbp]
\caption{List of the typical cosmological evolution behaviors of the bimetric massive gravity at $z \gg 1$.}
\vskip 0.2in
\label{tab:1}
\begin{tabular}{|c||c|c|c|} \hline
Parameters & $b/a \propto$ & $\rho_{DE} \propto$ & $g_{\mu \nu}$ and $h_{\mu \nu}$ coupling
\\ \hline \hline
$\beta_1 \neq 0$, $\beta_{2,3,4} = 0$ & $ H^{-2}$ & $ H^{-2} $ & Weak
\\ \hline
$\beta_2 \neq 0$, $\beta_{1,3,4} = 0$ & No solution & - & -
\\ \hline
$\beta_3 \neq 0$, $\beta_{1,2,4} = 0$ & $  H^2$ & $ H^4 $ & Strong
\\ \hline
$\beta_4 \neq 0$, $\beta_{1,2,3} = 0$ & $  H $ & $ H^3 $ & Strong
\\ \hline
\end{tabular}
\end{table}

As discussed in Refs.~\cite{Akrami2013, Koenning20141}, the $e_0\left(\mathbb{X}\right)$ term behaves as a trivial cosmological constant,
while the bimetric massive gravity is consistent with observation at the background level if $\beta_1 \neq 0$.
However, it is known that most of the modified gravity theories are hard to fit the CMB and large scale structure observations without
the $\Lambda$CDM limit, so that we will concentrate on our study with a non-zero value of $\beta_0$.
To simplify our discussion, we investigate the nontrivial case with only two free parameters, $\beta_0$ and $\beta_1$, which can recover the $\Lambda$CDM mode and the model described in Refs.~\cite{Akrami2013, Koenning20141} when the suitable $\beta_0$ and $\beta_1$ are chosen.
In addition, from Eqs.~\eqref{eq:eom1} and \eqref{eq:eom3}, if $\beta_1 = 0$, we see that there is no solution for $\beta_2 \neq 0$, while non-zero values of $\beta_3$ and $\beta_4$ yield too large abundance of the dark energy density in the early universe.
The behaviors of the cosmological evolutions with different one-parameter branches are listed in Table.~\ref{tab:1}.
Note that from the table, we see that the strong-coupling case would occur when $b/a$ is large for the non-zero parameters of $\beta_{3,4}$.
As a result,
we take $\beta_2 = \beta_3 = \beta_4 =0$ to be our minimal choice of the bimetric massive gravity theory.
In this scenario, Eqs.~(\ref{eq:eom3}) and (\ref{eq:eom4}) can be reduced to
\begin{eqnarray}
\label{eq:hboa}
\frac{b}{a} = \frac{ \beta_1}{3 H^2} \, \qquad \mathrm{and} \qquad H_b \equiv \frac{H_f}{H} = 1 - 2 \frac{\dot{H}}{H^2} \,,
\end{eqnarray}
respectively.

\begin{figure}
\centering
\includegraphics[width=0.48 \linewidth]{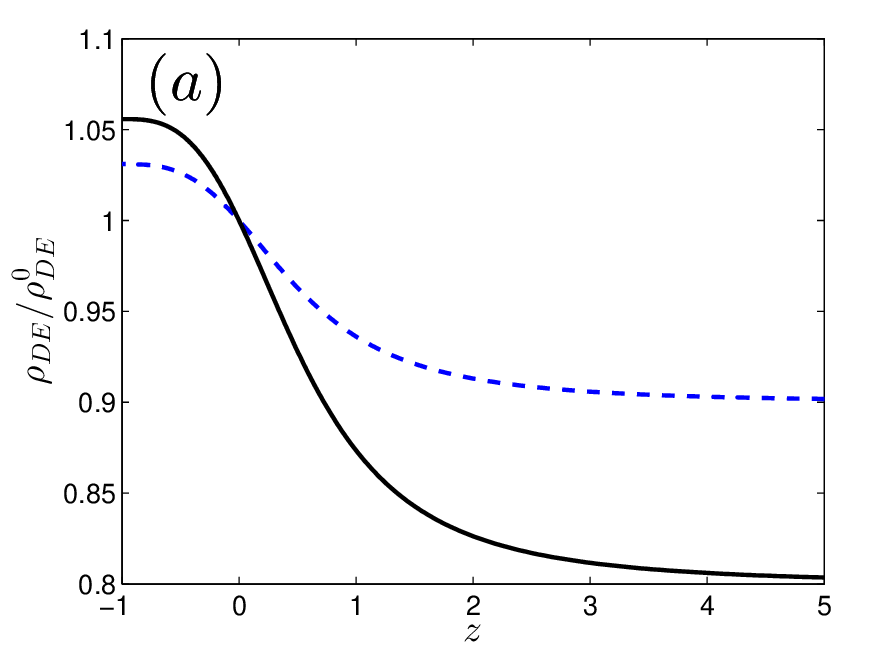}
\includegraphics[width=0.48 \linewidth]{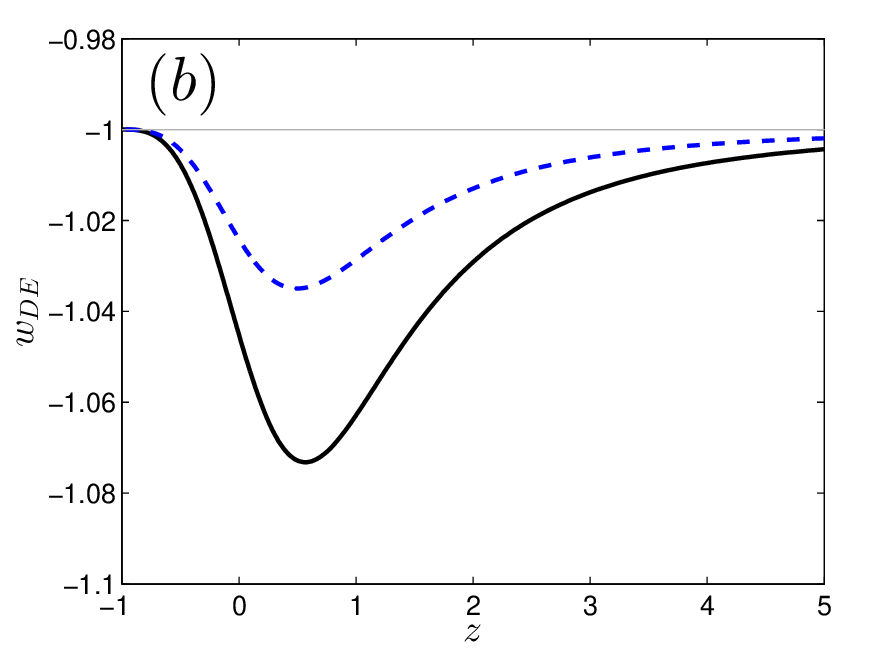}
\caption{Evolutions of (a) $\rho_{DE}$ and (b) $w_{DE}$ as functions of the redshift $z$ with $(\bar{\beta}_0, \bar{\beta}_1) = (0.8, 0.2)$ (black solid line) and $(0.9, 0.1)$ (blue dashed line), where the boundary conditions of $\Omega_m=0.26$ and $\Omega_r=8.4\times 10^{-5}$ are used.}
\label{fg:1}
\end{figure}

We emphasize that in order to avoid the instability~\cite{Crisostomi2012}, there is no mirror dark matter sector minimally coupled to the second metric in this work. 
Consequently, in terms of  Eqs.~(\ref{eq:eom1}) and (\ref{eq:eom2}),
the effective energy density and pressure can be defined by
\begin{eqnarray}
\label{eq:rhode}
&& \rho_{DE} = \beta_0 + 3 \beta_1 \frac{b}{a} = \rho_{DE}^{(0)} \left( \bar{\beta}_0 + \bar{\beta}_1 \frac{H_0^2}{H^2} \right) \,, \\
\label{eq:Pde}
&& P_{DE} = - \beta_0 - \beta_1 \left( 2 \frac{b}{a} + \frac{\dot{b}}{\dot{a}} \right) = \rho_{DE}^{(0)} \left[ -\bar{\beta}_0 + \bar{\beta}_1 \frac{H_0^2}{H^2} \left( \frac{2\dot{H}}{3H^2} - 1 \right) \right]  \,,
\end{eqnarray}
 which satisfy the continuity equation, $\dot{\rho}_{DE} + 3 H \left( \rho_{DE} + P_{DE} \right) = 0$, where we have redefined
\begin{eqnarray}
\label{eq:modelpar}
\bar{\beta}_0 = \frac{ \beta_0 }{ \rho_{DE}^{(0)}} \quad \mathrm{and} \quad \bar{\beta}_1 = \frac{\beta_1^2 }{ H_0^2 \rho_{DE}^{(0)}} \,,
\end{eqnarray}
with $\bar{\beta}_0 + \bar{\beta}_1 = 1$ and $\rho_{DE}^{(0)}$ the effective dark energy density at present. From Eqs.~(\ref{eq:rhode}) and (\ref{eq:Pde}), one can observe that $e_0 \left(\mathbb{X}\right)$ with the free parameter $\beta_0$ in the action plays the role of the cosmological constant, while $e_1 \left(\mathbb{X}\right)$ with $\beta_1$ behaves as the inverse quadratic of the Hubble parameter.
In Fig.~\ref{fg:1}, we illustrate the evolutions of $\rho_{DE}$ and equation of state (EoS), $w_{DE} = P_{DE} / \rho_{DE}$, as functions of the redshift $z$.
The effective dark energy density grows in the evolution of the universe and reaches a de-Sitter solution, $H \rightarrow H_{de} = const.$, in the future.
As shown in Fig.~\ref{fg:1}b, $w_{DE}$ always stays in the phantom phase of $w_{DE} < -1$ in our minimal scenario.
We note  that  this result has been already found in Ref.~\cite{Koenning20141}.

\section{Linear Perturbation Theory}
\label{sec:density-pert}

 From the generic linear perturbations in the $k$-space, we have
\begin{eqnarray}
\label{eq:pertgf}
g_{\mu\nu} = \bar{g}_{\mu\nu} + h_{\mu\nu} \,, \qquad
f_{\mu\nu} = \bar{f}_{\mu\nu} + h^f_{\mu\nu}\,,
\end{eqnarray}
where $\bar{g}_{\mu\nu}$ and $\bar{h}_{\mu\nu}$ are the background metrics,
while
$h_{\mu\nu}$ and $h^f_{\mu\nu}$ are the linear perturbations in Eqs.~(\ref{eq:g_bg}) and (\ref{eq:f_bg}), respectively.
In general, there are four scalar modes in each metric, i.e.,
\begin{eqnarray}
\label{eq:pertg}
&& h_{00} = -2 \Phi \,, \quad h_{0i} = h_{i0} = 2 a^2 \partial_i B \,, \quad h_{ij} = 2 a^2 \left[ \frac{k_i k_j}{2 k^2} h + \left( 3 \frac{k_i k_j}{k^2} - \delta_{ij} \right) \eta \right] \,, \\
\label{eq:pertf}
&& h^f_{00} = -2 \frac{\dot{b}^2}{\dot{a}^2} \Phi_f \,, \quad h^f_{0i} = h^f_{i0} = 2 b^2 \partial_i B_f \,, \quad h^f_{ij} = 2 b^2 \left[ \frac{k_i k_j}{2 k^2} h_f + \left( 3 \frac{k_i k_j}{k^2} - \delta_{ij} \right) \eta_f \right] \,,
\end{eqnarray}
where $k^2 = k_i k^i$ and $i=1, 2, 3$.
Under a gauge transformation,
\begin{eqnarray}
x^{\prime \mu} = x^{\mu} + \epsilon^{\mu} \,,
\end{eqnarray}
with $\epsilon^{\mu}$ at the same order of $h_{\mu\nu}$ and $h^f_{\mu\nu}$, we can eliminate two of these scalars.
In order to compare the results with the observations, we choose the conventional synchronous gauge with two scalars, ($h$ and $\eta$), in the ordinary metric~\cite{Ma:1995ey}, and keep four scalars, ($h_f$, $\eta_f$, $\Phi_f$ and $B_f$), in the new one.

Substituting Eqs.~(\ref{eq:pertg}) and (\ref{eq:pertf}) with $\Phi = B = 0$ into Eqs.~(\ref{eq:fieldeq1}) and (\ref{eq:fieldeq2}), we have
\begin{eqnarray}
\label{eq:soleqg1}
&& h^{\prime \prime} + \left( 2 + \frac{H^{\prime}}{H} \right) h^{\prime} = \frac{\beta_1^2}{6H^4} \left( h - h_f - \frac{3\Phi_f}{H_b} \right) - 3 \frac{\left( 1+3w_M \right) \rho_M}{\rho_M+\rho_{DE}} \delta_M \,, \\
\label{eq:soleqg2}
&& \frac{k^2 \eta^{\prime}}{a} = \frac{(\bar{\rho}_M + \bar{P}_M) \theta_M}{2 H} + \frac{\beta_1^2}{3H^3} \frac{k^2 B_f}{1+ H_b} \,,
\end{eqnarray}
and
\begin{eqnarray}
\label{eq:soleqf1}
&& h_f^{\prime} + 6 \eta_f^{\prime} =  \frac{3 H_b}{2} \left( h - h_f \right) + 2 H_b \frac{k^2 \eta_f}{a^2 H^2} - 4 \frac{k^2}{a^2 H^2} \bar{B}_f - \frac{18 H_b^3}{1+H_b} \bar{B}_f \,, \\
\label{eq:soleqf2}
&& \frac{3 H_b^2}{1+H_b} \bar{B}_f^{\prime} - \left( H_b - 1 - \frac{H^{\prime}}{H} + \frac{3}{2} H_b^2 \right)  \frac{\eta_f^{\prime}}{H_b} \nonumber \\
&& \qquad = 3 \left( \eta_f - \eta \right) - \left( 2 + 2 \frac{H^{\prime}}{H} + H_b - \frac{3}{2} H_b^2 + \frac{ \left( 2+ H_b \right) H_b^{\prime} }{\left( 1+H_b \right) H_b} \right)  \left( \frac{3 H_b^2}{1+H_b} \bar{B}_f \right) \,, \\
\label{eq:soleqf3}
&&h_f^{\prime} + \frac{12 H_b^2}{1+H_b} \bar{B}_f^{\prime} = h^{\prime} + 12 \left( \eta_f - \eta \right) + \left( \frac{H^{\prime}}{H} + 2 H_b -1 \right) \left( h-h_f \right) + \frac{4 k^2 \eta_f}{3a^2 H^2} \left( H_b - 1 - \frac{H^{\prime}}{H} \right) \nonumber \\
&& \qquad - \left( \frac{12 H_b^2}{1+H_b} \bar{B}_f \right) \left[ \frac{k^2 H_b}{3 a^2 H^2} + 2 H_b +  \frac{(2+H_b)H_b^{\prime}}{(1+H_b)H_b} + 1 + \frac{H^{\prime}}{H} \right] \,, \\
\label{eq:soleqf4}
&& \Phi_f = - \left( \frac{3 H_b^2}{1+H_b} \bar{B}_f + \frac{\eta_f^{\prime}}{H_b} \right) \,,
\end{eqnarray}
where the prime denotes the derivative of e-folding, i.e. ``$\prime$'' $= d/dN = d/d \ln a$, $\bar{B}_f = a H B_f$, $\tilde{H} \equiv H / H_b - \dot{H}_b / 2 H^3_b$, $\delta T^0_0 = \delta \rho_M = \rho_M \delta_M$, $\delta T^0_i = - \delta T^i_0 = (\rho_M + P_\ell) v^i_M$, $\delta
 T^i_j = \delta P_M \delta^i_j$ and $\theta_M \equiv \partial_i v^i_M$.
In addition, from the conservation equation $\nabla^{\mu} T^M_{\mu \nu} = 0$, one gets
\begin{eqnarray}
\label{eq:continuity1}
&& \delta_M^{\prime} = - \left( 1+w_M \right) \left( \theta_M + \frac{h^{\prime}}{2} \right) - 3 \left( \frac{\delta P_M}{\delta \rho_M} -w_M \right) \delta_M \,, \\
\label{eq:continuity2}
&& \theta_M^{\prime} = -\left( 1-3w_M \right) \theta_M - \frac{w_M^{\prime}}{1+w_M} \theta + \frac{\delta P_M / \delta \rho_M}{1+w_M} \frac{k^2 \delta_M}{H} \,.
\end{eqnarray}
From Eqs.~(\ref{eq:soleqg1}), (\ref{eq:soleqg2}) (\ref{eq:continuity1}) and (\ref{eq:continuity2}), it is easy to check that the bimetric massive gravity theory with $(\bar{\beta}_0, \bar{\beta}_1) = (1, 0)$ is reduced to the $\Lambda$CDM limit at not only the background evolution level, but also the linear perturbation one.

\section{Cosmological Evolution of Matter and Scalar Perturbations}
\label{sec:mpk}

The Hubble radius $d_H \equiv H^{-1}$ expands during the evolution of the universe.
More and more density perturbation modes of $\delta_k$  enter the horizon with the wavenumber $k$.
By using $\Omega_m=0.26$, $\Omega_r=8.4 \times 10^{-5}$ and $H_0=70 km/s \cdot Mpc$, $\delta_k$ reaches the horizon at the redshifts $z_k \simeq 32$ and $8 \times 10^{4}$ with $k=10^{-3}$ and $0.25 [h/Mpc]$, respectively.
The matter power spectrum $P(k) \sim \langle \delta_m^2 (k) \rangle$ with the BBKS transfer function can be recovered within $10\%$ accuracy by taking the initial density perturbation to be the scale invariance when all modes are located at the super-horizon scale, i.e., $\delta_m = 3 \delta_r /4 \propto k^{n_s/2}$.
We emphasize that although the the accurate values of $\Omega_m, \Omega_r$ and $H_0$ in the bimetric massive gravity theories
are needed to be determined by the observational data, they should not significantly deviate from those in the $\Lambda$CDM theory.
Therefore, it is reasonable to choose the evolution of $\delta_m$ with the initial scale factor of $\ln a_i=-18$.

\begin{figure}
\centering
\includegraphics[width=0.48 \linewidth]{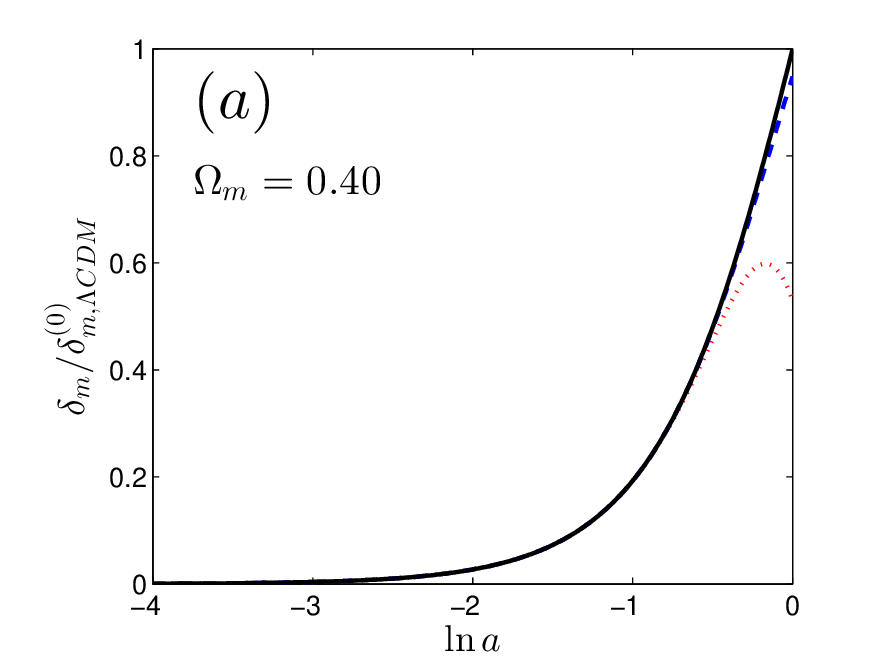}
\includegraphics[width=0.48 \linewidth]{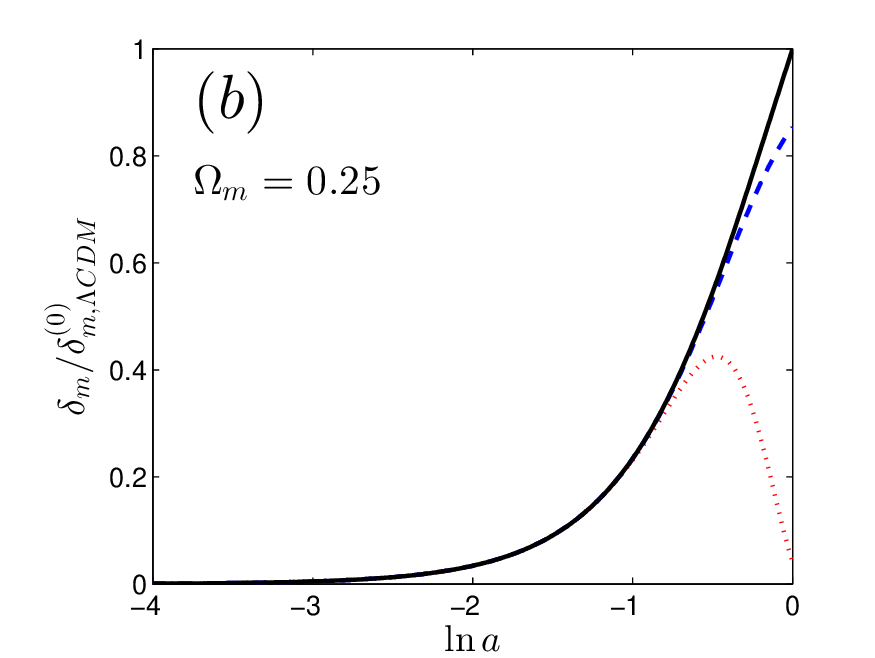}
\caption{ Evolutions of the matter density perturbation $\delta_m$, normalized by that in the $\Lambda$CDM limit at $z=0$, as a function of $N = \ln a$ by fixing $k=0.1 [h/Mpc]$  with $(\bar{\beta}_0, \bar{\beta}_1) = (1, 0)$ (black solid line), $(0.99, 0.01)$ (blue dashed line) and $(0.9, 0.1)$ (red dotted line), respectively, where $\Omega_r = 8 \times 10^{-5}$ is used.}
\label{fg:2}
\end{figure}

\begin{figure}
\centering
\includegraphics[width=0.48 \linewidth]{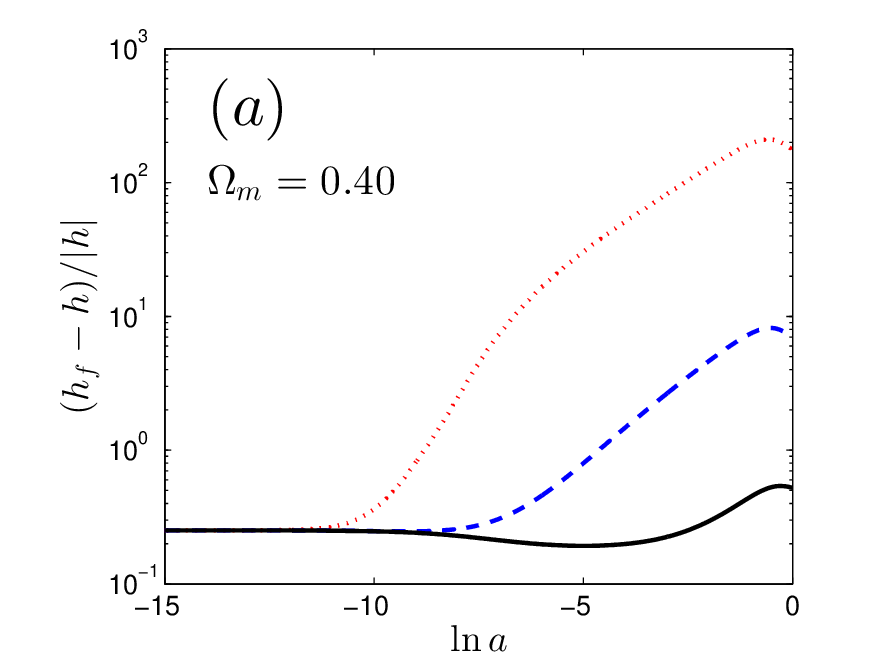}
\includegraphics[width=0.48 \linewidth]{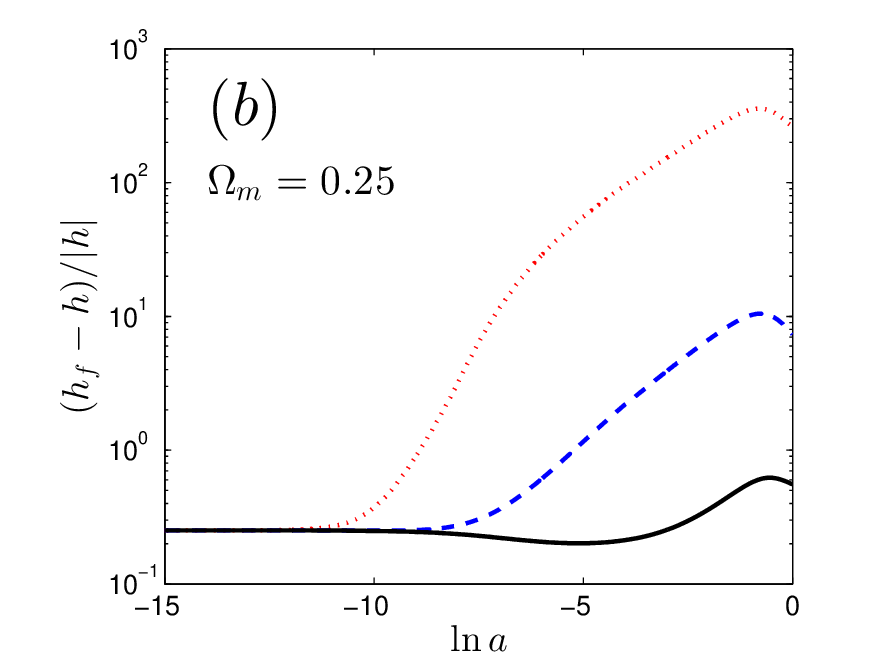}
\caption{ Evolutions of $(h_f-h)/|h|$ with $(\bar{\beta}_0, \bar{\beta}_1) = (0.99, 0.01)$, and $k= 10^{-3}[h/Mpc]$ (black solid line), $10^{-2}[h/Mpc]$ (blue dashed line) and $0.1 [h/Mpc]$ (red dotted line), respectively, where the boundary conditions are taken to be the same as those in Fig.~\ref{fg:2}.}
\label{fg:4}
\end{figure}

By using Eqs.~(\ref{eq:soleqg1}) - (\ref{eq:continuity2}), the evolutions of $\delta_M$ and $\theta_M$ as well as the linear perturbation scalars, such as $h$, $\eta$, $h_f$, $\eta_f$, $B_f$ and $\Phi_f$, can be solved.
In Fig.~\ref{fg:2}, we present the evolution of the normalized matter density perturbation, $\delta_m/\delta_{m,\Lambda CDM}^{(0)}$, as a function of the e-folding $N \equiv \ln a$ by fixing $k=0.1 [h/Mpc]$ with $(\bar{\beta}_0, \bar{\beta}_1) = (1, 0)$ (black solid line), $(0.99, 0.01)$ (blue dashed line), and $(0.9, 0.1)$ (red dotted line),
 where $\delta_{m,\Lambda CDM}^{(0)}$ is the value in the $\Lambda$CDM limit at $z=0$ (a) and (b) correspond to  $\Omega_m = 0.4$ and  $0.25$, respectively.
When $h_f+3\Phi_f/H_b > h$ in Eq.~(\ref{eq:soleqg1}), the growth of the matter density perturbation is smaller than that in the $\Lambda$CDM case ($\bar{\beta}_1 = 0$).
As a result, $\delta_m$ in the bimetric massive gravity theory is suppressed compared to that in the $\Lambda$CDM model.
Clearly, a larger dark energy density leads to a bigger suppression $\delta_m$.
As a result, it is reasonable to conclude that 
the  Vainshtein mechanism becomes effective when    $k \gtrsim O(0.1-1) ~ [h/Mpc]$.

In Fig.~\ref{fg:4}, the difference of the scalars, $(h_f-h)/|h|$, is represented as a function of $N$ with $(\bar{\beta}_0, \bar{\beta}_1) = (0.99, 0.01)$,  and $k= 10^{-3}[h/Mpc]$ (black solid line), $10^{-2}[h/Mpc]$ (blue dashed line), and $0.1 [h/Mpc]$ (red dotted line),
 where   (a) $\Omega_m = 0.4$ and (b) $\Omega_m = 0.25$, respectively.
At the super-horizon scale, we see that $h_f \gtrsim h$ in the evolution of the universe, while the detailed calculations in Eqs.~(\ref{eq:A.radsuper}) and (\ref{eq:A.matsuper}) show that $0.75h \simeq h_f < 0$ in both radiation and matter dominated epochs.
When the $k$-mode is in the horizon, $h_f$ sharply increases with
\begin{eqnarray}
\label{eq:solhf}
h_f \propto a^{\bar{\lambda}+c} \quad \mathrm{and} \quad \eta_f \propto a^{\bar{\lambda}} \,,
\end{eqnarray} 
where $c=2$ and $1$ at the radiation and matter dominated epochs, respectively.
In Eqs.~(\ref{eq:A.radsub}) and (\ref{eq:A.matsub}), we find that $\bar{\lambda} \sim 0.5$ before the dark energy dominated era.
The larger $k$ is, the earlier the mode enters the horizon, resulting in a more significant enhancement of $h_f$.
Although the bimetric massive gravity theory restrains the growth of the matter density perturbation, the suppression effect of $\delta_m$ would be negligible by $H^{-4}$ in the early time (see also Eq.~(\ref{eq:soleqg1})) until
\begin{eqnarray}
\label{eq:subhorizon}
\left| \frac{\beta_1^2 h_f}{H^4} \right| > \delta_M \,.
\end{eqnarray}
Clearly, such a suppression depends on both the wavenumber $k$ and model parameter $\beta_1$.

\section{Ghost and instability problems and Matter Power Spectrum}
\label{sec:stability}

\begin{figure}
\centering
\includegraphics[width=0.48 \linewidth]{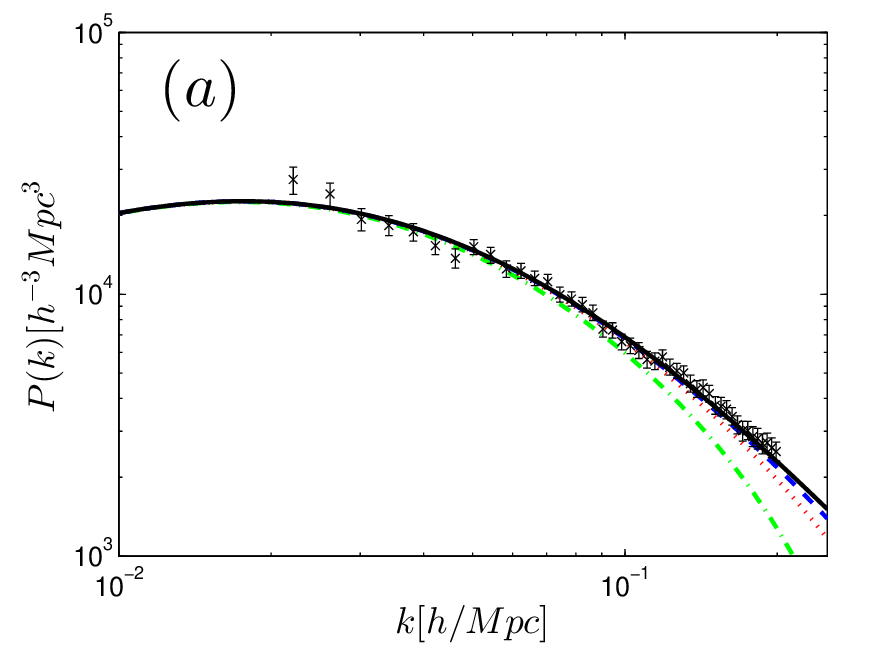}
\includegraphics[width=0.48 \linewidth]{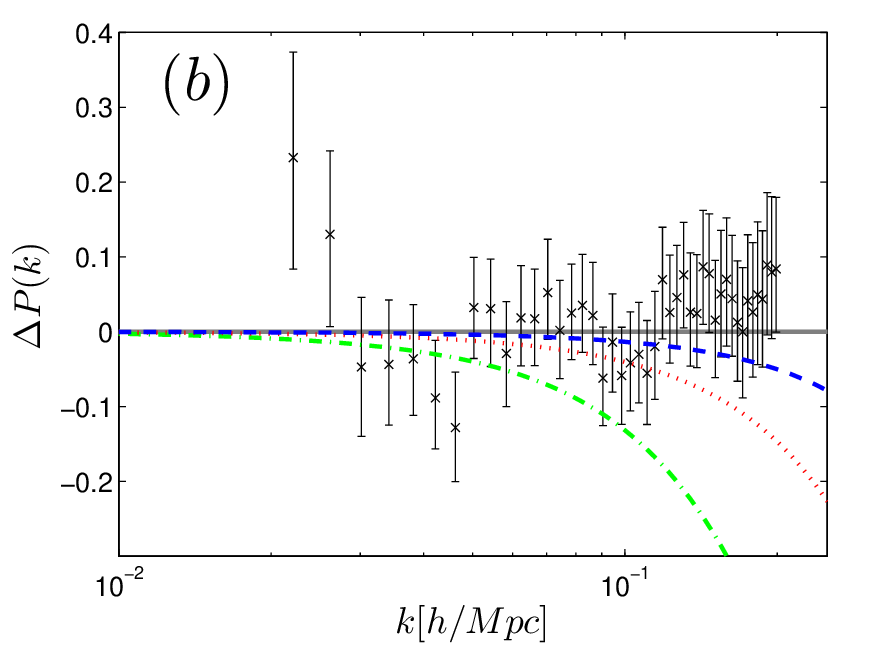}
\caption{(a) The matter power spectrum $P(k)$ and (b) $\Delta P(k) = (P - P_{\Lambda \mathrm{CDM}})/P_{\Lambda \mathrm{CDM}}$ as functions of the wavenumber $k$ with $\sum m_{\nu} = 0.2$~eV and $\bar{\beta}_1 = 0$ (black solid line), $10^{-3}$ (blue dashed line), $3 \times 10^{-3}$ (red dotted line) and $10^{-2}$ (green dash-dotted line), together with the data points come from the SDSS LRG DR7.
Note that $P_{\Lambda \mathrm{CDM}}$ corresponds to the case with $\bar{\beta}_1 = 0$, and the boundary conditions are taken to be the same as those in Fig.~\ref{fg:1}.}
\label{fg:3}
\end{figure}

It is known that the bimetric massive gravity suffers from the Higuchi ghost and instability problems~\cite{
Higuchi:1986py,Higuchi:1989gz,Crisostomi2012, Solomon2014, Koenning2014, Lagos2014, Comelli2014prd, Koenning2014prd2, Berg2012,Lagos2016JCAP, Koenning20141, Akrami2014, Felice2014, Akrami20131, Rham2015, Strauss2012, Volkov2012,Katsuragawa2014}, due to the negative squared mass of graviton and the divergence under the linear density perturbation, respectively.
In order to avoid the ghost problem, the condition of $(b/a)^{\prime} \geq 0$ is required, which can be derived from Ref.~\cite{Koenning2015}.
From Eq.~\eqref{eq:hboa}, we see that $(b/a)^{\prime} \propto - \dot{H}/H^4$, which is larger than or equal to zero with the equal sign at $H=H_{de}$ in the late-time dark energy dominated epoch. Clearly, our minimum case of
the bimetric massive gravity theory is ghost-free in the cosmological evolution.
On the other hand, it has been checked in Ref.~\cite{Koenning2015} that the scalar perturbations
with $b^2/a^2<1/3$ and $(b/a)^{\prime} \simeq a (b/a)^{\prime 2} / b$ in the radiation and dark energy dominated epochs
 are stable in both early and late-time of universe, respectively, but the stability condition
is broken in the matter dominated epoch with $\beta_i = 0$ ($i=2, 3$ and $4$) as obtained in Ref.~\cite{Koenning2014}.
As shown in Fig.~\ref{fg:4}, our numerical result confirms such a divergence property on the growth of the linear scalar perturbation for the new metric, $h_f$, in the sub-horizon scale.
The sharply increasing $h_f$ leads to the suppression of the cosmological observables, $\delta_m$, $h$ and $\eta$.
Fortunately, the suppression strength, which can be estimated from Eqs.~\eqref{eq:soleqg1} and \eqref{eq:soleqg2}, is proportional to $\beta_1^2 \sim \bar{\beta}_1$.
As a result, when $\bar{\beta}_1$ is small enough, the deviation of $\delta_m$ in the bimetric massive gravity theory from that in the $\Lambda$CDM model can be restricted.
By comparing the deviation of the matter power spectrum, $P(k) \propto |\delta_m|^2$, to the large scale structure observations, we can give an upper bound for $\bar{\beta}_1$.
In Fig.~\ref{fg:3}, we illustrate $P(k)$ as a function of $k$ with a massive neutrino of $m_{\nu} = 0.2$~eV and $\bar{\beta}_1 = 0$ (black solid line), $10^{-3}$ (blue dashed line), $3 \times 10^{-3}$ (red dotted line) and $10^{-2}$ (green dash-dotted line), respectively, where the data points come from the SDSS LRG DR7~\cite{Abazajian:2008wr}.
Combining Eqs.~\eqref{eq:continuity1}, \eqref{eq:solhf} and \eqref{eq:A.guess3} with $h_f < 0$ in the matter dominated era at the scale 
inside the deep horizon, Eq.~\eqref{eq:soleqg1} reduces to
\begin{eqnarray}
\delta_m^{\prime \prime} + \left( 2 + \frac{H^{\prime}}{H} \right) \delta_m^{\prime} - \frac{\beta_1^2}{12 H^4} h_f - \frac{3}{2} \delta_m = 0 \,,
\end{eqnarray}
where the third term is positive, reducing the growth of $\delta_m$, which leads to the suppression of $P(k)$.
Comparing the numerical calculation with the data points up to $k \simeq 0.1 h/Mpc$, we can make a statement that the allowed window is $\bar{\beta}_1 \lesssim \mathcal{O}(10^{-2})$ in the bimetric massive gravity theory.
 Note that (i) the perturbation theory breaks when $h^{f}_{\mu \nu}$ is large enough, occurring at $k \sim 0.1 [h/Mpc]$, indicating that the linear perturbation result is applicable only up to $k \sim 0.1$, and (ii)
our upper bound is consistent with the conclusion that a positive $\beta_1$ is needed for avoiding instability in Ref.~\cite{Lagos2014}.
We note that $\bar{\beta}_1$ keeps positive even if the stability condition is relaxed with $\beta_1 < 0$.
Thus, the background evolution analysis is still valid along with the suppressed  $P(k)$  in this gravitational theory.

Additionally, one has
\begin{eqnarray}
\label{eq:stabde}
\frac{ ( b/a )^{\prime \prime}}{b/a} = 9(1+w_t)^2+3w_t^{\prime} \lesssim \left( \frac{ ( b/a )^{\prime}}{b/a} \right)^2 = 9(1+w_t)^2 \,.
\end{eqnarray}
which satisfies the stability condition in Ref.~\cite{Koenning2015} in the dark energy dominated era, where we have used $w_t = (P_M + P_{DE})/(\rho_M + \rho_{DE})$ and $w_t^{\prime} \lesssim 0$.
Consequently, although under the linear scalar perturbations the observables are unstable in the matter dominated epoch, they become stabilized again when dark energy dominates the universe.

\section{Conclusions}
\label{sec:conclusion}

We have studied the matter density perturbation $\delta_m$ and matter power spectrum $P(k)$ in the bimetric massive gravity theory for the minimal scenario with $\beta_2=\beta_3=\beta_4=0$.
In this scenario, the zero values for $\beta_{3,4}$
  are used to avoid the strong coupling between $g_{\mu \nu}$ and $h_{\mu \nu}$, while  $\beta_1 > 0$ is taken to evade 
  the stability problem.
In this approach, $\beta_0 e_0 \left(\mathbb{X} \right)$ in the action plays the role of the cosmological constant in both background and linear perturbation levels, while $\beta_1 e_n \left(\mathbb{X} \right)$ behaves as a inverse quadratic of $H$.
This result confirms the analysis in Ref.~\cite{Koenning20141} that the effective dark energy EoS in the theory is always at the phantom phase, i.e., $w_{DE}<-1$.

By taking the synchronous gauge in $g_{\mu\nu}$, we have calculated the evolution of the matter density perturbation in our minimal case of the bimetric massive gravity theory.
Through the analytical discussion, the growth of the scalars in the new metric is independent on the choices of $\beta_0$ and $\beta_1$.
At the scale outside the horizon, $k^2/a^2 \ll H^2$, we have the relations, $h_f \sim h \gg \eta_f \sim \Phi_f \gg \bar{B}_f$.
When the scale enters the horizon, $k^2/a^2 \gg H^2$, the growths of $\eta_f$, $\Phi_f$ and $\bar{B}_f$ are suspended, and $\lvert h_f \rvert$ sharply increases,  reducing the growth of the matter density fluctuation and causing the suppression of the matter power spectrum $P(k)$
in both $\beta_1 > 0$ and $\beta_1 < 0$ cases.
This behavior is the same as the DGP model~\cite{Falck:2014jwa}, but opposite to the cubic Galileon one~\cite{Li:2013tda}.
This effect depends on both the scale $k$ and parameter set $(\beta_0, \beta_1)$.

Even if the linear scalar perturbations in the new metric diverge in the matter dominated epoch, $\delta_m$ can still fit to the cosmological observations with a small enough $\bar{\beta}_1$.
In addition, the perturbation theory breaks at $k \sim 0.1 [h/Mpc]$, and the linear perturbation result is not applicable to the larger $k$.
Comparing our numerical results of $P(k)$ with the data, we claim that the allowed model parameter is $\bar{\beta}_1 \lesssim \mathcal{O}(10^{-2})$ at the linear perturbation level.
For completeness, we have to explore the non-linear perturbation theory on the new metric, $f_{\mu \nu}$,
while the exact preferred value of $\beta_1$ should be further investigated by comparing the theoretical predictions with all of the observational data, such as those from the type-Ia supernova, CMB and weak lensing, which will be presented elsewhere.

\section*{Acknowledgments}
This work was partially supported by the National Natural Science Foundation of China under Grant No. 11505004, National Center for Theoretical Sciences and MoST (MoST-104-2112-M-007-003-MY3), and the Anhui Provincial Natural Science Foundation of China under Grant No. 1508085QA17. We thank Yashar Akrami and Marco Crisostomi for useful communications.

\appendix

\section{Growths of Scalar Modes in Bimetric Massive Gravity}

We present the detailed calculation of the evolution equations for the scalar perturbations in the bimetric massive gravity theory, including two scalar perturbations, $h$ and $\eta$, in the ordinary metric and four scalar perturbations, $h_f$, $\eta_f$, $B_f$ and $\Phi_f$, in the new metric.
From Eqs.~(\ref{eq:soleqf1}) - (\ref{eq:soleqf3}), we can observe that the evolutions of the scalar perturbations in the new metric is independent on the model parameter $\beta_1$, so that $\beta_1 \rightarrow 0$ is taken in our discussion of  the asymptotic behavior in the radiation and matter dominated epochs in this section.

\subsection{Scalar perturbations at the super-horizon scale, $k^2 \ll a^2 H^2$}

At the super-horizon scale, Eqs.~(\ref{eq:soleqf1}) - (\ref{eq:soleqf3}) are reduced to
\begin{eqnarray}
&& h_f^{\prime} + 6 \eta_f^{\prime} \simeq \frac{3(4+3w_M)}{2} \left( h-h_f \right) - \frac{18 (4+3w_M)^3}{5+3w_M} \bar{B}_f \,, \\
&& \frac{3(4+3w_M)^2}{5+3w_M} \bar{B}_f^{\prime} -  \frac{( 57 + 81 w_M + 27w_M^2)}{2(4+3w_M)} \eta_f^{\prime} \simeq 3 \left( \eta_f - \eta \right) \nonumber \\
&& \qquad \qquad \qquad \qquad \qquad \qquad \qquad \qquad + \frac{9(4+3w_M)^2(14+24w_M+9w_M^2)}{2(5+3w_M)} \bar{B}_f \,, \\
&& h_f^{\prime} + \frac{12 (4+3w_M)^2}{5+3w_M}\bar{B}_f^{\prime} \simeq h^{\prime} + 12 \left( \eta_f - \eta \right) + \frac{(11+9w_M)}{2} \left( h-h_f \right) - 18(4+3w_M)^2 B_f \,, \qquad
\end{eqnarray}
where we have used $w_M=const.$, $\dot{H}/H^2 = -3(1+w_M)/2$ and $H_b=4+3w_M$, and redefined $\bar{B}_f = a H B_f$, while the prime denotes the derivative of e-folding, i.e. ``$\prime$'' $= d/d \ln a$.

It is known that $\delta_M$ and $ h \propto a^{\lambda}$ in the radiation ($w_M=1/3$, $\lambda=2$) and matter ($w_M=0$, $\lambda=1$) dominated epoches, respectively.
Substituting the relations,
\begin{eqnarray}
\label{eq:A.guess}
h_f \propto a^{\lambda_h} \,, \quad \eta_f \propto a^{\lambda_{\eta}} \quad \mathrm{and} \quad B_f \propto a^{\lambda_B} \,,
\end{eqnarray}
into Eqs.~(\ref{eq:soleqf1}) - (\ref{eq:soleqf3}), the only possible solution is
\begin{eqnarray}
\lambda_h=\lambda_\eta=\lambda_B=\lambda \,.
\end{eqnarray}
Combining the equations above, the scalar perturbations are found to b explicitly functions of $h$ and $\eta$, given by
\begin{eqnarray}
\label{eq:A.radsuper}
h_f \sim 0.748 h \,, \quad \eta_f \sim -0.057 h + 0.405 \eta \quad \mathrm{and} \quad B_f \sim 2.87 \times 10^{-3} h -0.013 \eta
\end{eqnarray}
and
\begin{eqnarray}
\label{eq:A.matsuper}
h_f \sim 0.822 h \,, \quad \eta_f \sim -0.040 h + 0.585 \eta \quad \mathrm{and} \quad B_f \sim 2.08 \times 10^{-3} h -0.015 \eta
\end{eqnarray}
in the radiation and matter dominated eras, respectively.

\subsection{ Scalar perturbations at the sub-horizon scale, $k^2 \gg a^2 H^2$}

When the $k$-mode enters the horizon, the $k^2/a^2H^2$ terms play the most important role in the cosmological evolution.
Substituting Eq.~(\ref{eq:A.guess}) into Eqs.~(\ref{eq:soleqf1}) - (\ref{eq:soleqf3}), the growth powers are deduced as
\begin{eqnarray}
\label{eq:A.guess2}
\lambda = \lambda_\eta = \lambda_B = \lambda_h - c \,,
\end{eqnarray}
where $c=2$ and $1$ for the radiation ($w_M=1/3$) and matter ($w_M=0$) dominated eras, respectively.
Thus, we have
\begin{eqnarray}
\label{eq:A.guess3}
h_f \sim \frac{k^2\eta_f}{a^2 H^2} \quad \mathrm{and} \quad \frac{k^2\bar{B}_f}{a^2 H^2} \,,
\end{eqnarray}
which allow us to take that $\bar{B}_f = \mathcal{C} \eta_f \ll h_f$ at the scale deep inside the horizon.
Finally, the values of $\mathcal{C}$ and $\lambda$ can be solved by substituting Eq.~(\ref{eq:A.guess2}) into Eqs.~(\ref{eq:soleqf1}) - (\ref{eq:soleqf3}), given by
\begin{eqnarray}
\label{eq:A.radsub}
w_M=\frac{1}{3}: \quad && \lambda \simeq 0.577 \,, \quad \mathrm{and} \quad \mathcal{C} \simeq -0.019 \,, \\
\label{eq:A.matsub}
w_M=0: \quad && \lambda \simeq 0.432 \,, \quad \mathrm{and} \quad \mathcal{C} \simeq -0.031 \,,
\end{eqnarray}
which match our numerical calculations of $\lambda \sim 0.58$ for $w_M=1/3$ and $\lambda \sim 0.43$ for $w_M=0$.
These results show a clear behavior that the growth of $h_f$ is enhanced with $h_f \propto a^{\lambda + c}$, but those of $\eta_f$ and $B_f$ are suppressed with $|\eta_f| \gg |\ \bar{B}_f| \propto a^{\lambda}$ in the matter dominated era.
As a result, we conclude that
\begin{eqnarray}
h_f \sim h
\end{eqnarray}
at the super-horizon scale, and
\begin{eqnarray}
\label{eq:A.approx}
|h_f| \gg |h| \gg |\eta_f| > |\bar{B}_f|
\end{eqnarray}
at the scale deep inside the horizon.


\begin{thebibliography}{99}

\bibitem{Perlmutter1999} S. Perlmutter,  et al., Astrophys. J.  {\bf 517}, 565 (1999).
\bibitem{Riess1998} A. G. Riess,   et al., Astron. J. {\bf 116}, 1009 (1998).
\bibitem{Tegmark2004} M. Tegmark, et al., Phys. Rev. D {\bf 69} 103501 (2004).
\bibitem{Eisenstein2005}D. J. Eisenstein, et al., Astron. J. {\bf 633}, 560 (2005).
\bibitem{Spergel} D. N. Spergel, et al., Astrophys. J. Suppl. {\bf 148}, 175 (2003);
                            D. N. Spergel, et al., Astrophys. J. Suppl. {\bf 170}, 377 (2007).

\bibitem{Dvali2000}D. Dvali, G. Gabadadze and M. Porrati, Phys. Lett. B {\bf 485},  208 (2000).

\bibitem{Fierz1939} M. Fierz and W. Pauli, Proc. Roy. Soc. Lond. A {\bf 173}, 211 (1939).

\bibitem{Dam1970} H. van Dam, M. Veltman, Nucl. Phys. B {\bf 22}, 397 (1970).
\bibitem{Zakharov1970} V. I. Zakharov,  J. Exp. Theor. Phys. Lett. {\bf 12}, 312 (1970);
                          V. I. Zakharov, Pisma Zh. Eksp. Teor. Fiz. {\bf 12}, 447 (1970).
\bibitem{Iwasaki1970} Y. Iwasaki, Phys. Rev. D {\bf 2}, 2255 (1970).

\bibitem{Vainshtein1972} A. I. Vainshtein, Phys. Lett. B {\bf 39}, 393 (1972).

\bibitem{Lue:2004rj}
  A.~Lue, R.~Scoccimarro and G.~D.~Starkman,
  Phys.\ Rev.\ D {\bf 69}, 124015 (2004)
  doi:10.1103/PhysRevD.69.124015
  [astro-ph/0401515].

\bibitem{Koyama:2007ih}
  K.~Koyama and F.~P.~Silva,
  Phys.\ Rev.\ D {\bf 75}, 084040 (2007)
  doi:10.1103/PhysRevD.75.084040
  [hep-th/0702169 [HEP-TH]].

\bibitem{Koyama:2009me}
  K.~Koyama, A.~Taruya and T.~Hiramatsu,
  Phys.\ Rev.\ D {\bf 79}, 123512 (2009)
  doi:10.1103/PhysRevD.79.123512
  [arXiv:0902.0618 [astro-ph.CO]].

\bibitem{Koyama:2013paa}
  K.~Koyama, G.~Niz and G.~Tasinato,
  Phys.\ Rev.\ D {\bf 88}, 021502 (2013)
  doi:10.1103/PhysRevD.88.021502
  [arXiv:1305.0279 [hep-th]].

\bibitem{Barreira:2014zza} 
  A.~Barreira, B.~Li, W.~A.~Hellwing, L.~Lombriser, C.~M.~Baugh and S.~Pascoli,
  JCAP {\bf 1404}, 029 (2014)
  doi:10.1088/1475-7516/2014/04/029
  [arXiv:1401.1497 [astro-ph.CO]].

\bibitem{Koyama:2015oma}
  K.~Koyama and J.~Sakstein,
  Phys.\ Rev.\ D {\bf 91}, 124066 (2015)
  doi:10.1103/PhysRevD.91.124066
  [arXiv:1502.06872 [astro-ph.CO]].

\bibitem{Koyama:2015vza}
  K.~Koyama,
  Rept.\ Prog.\ Phys.\  {\bf 79}, no. 4, 046902 (2016)
  doi:10.1088/0034-4885/79/4/046902
  [arXiv:1504.04623 [astro-ph.CO]].

\bibitem{Crisostomi:2017lbg}
  M.~Crisostomi and K.~Koyama,
  arXiv:1711.06661 [astro-ph.CO].


\bibitem{Boulware1972} D. G. Boulware and S. Deser, Phys. Rev. D {\bf 6}, 3368 (1972).

\bibitem{Rham2010} C. de Rham, and G. Gabadadze, Phys. Rev. D {\bf 82}, 044020 (2010) [arXiv:1007.0443].
\bibitem{Rham2011} C. de Rham, G. Gabadadze, and A. J. Tolley, Phys. Rev. Lett. {\bf 106}, 231101 (2011) [arXiv:1011.1232].

\bibitem{Hassan20123} S. F. Hassan, and R. A. Rosen, JHEP {\bf 04}, 123 (2012) [arXiv:1111.2070].

\bibitem{Amico2011} G. D’Amico, C. de Rham, S. L. Dubovsky, G. Gabadadze, and D. Pirtskhalava, and A. J. Tolley, Phys. Rev. D {\bf 84}, 124046 (2011) [arXiv:1108.5231].
\bibitem{Felice2012} A. De Felice, A. E. Gumrukcuoglu, and S. Mukohyama, Phys. Rev. Lett. {\bf 109}, 171101 (2012).

\bibitem{Rham2014} C. de Rham, Living Rev. Rel. {\bf 17}, 7 (2014).



\bibitem{Hassan20121} S. F. Hassan, and R. A. Rosen, JHEP {\bf 02}, 126 (2012) [arXiv:1109.3515].
\bibitem{Hassan20122} S. F. Hassan, R. A. Rosen, and A. Schmidt-May, JHEP {\bf 02}, 026 (2012) [arXiv:1109.3230].


\bibitem{Akrami2013} Y. Akrami, T. Koivisto, M. Sandstad, JHEP {\bf 03}, 099 (2013).
\bibitem{Volkov2013} M. Volkov, Class. Quant. Grav. {\bf 30}, 184009 (2013).

\bibitem{Higuchi:1986py}
  A.~Higuchi,
  Nucl.\ Phys.\ B {\bf 282}, 397 (1987).
  doi:10.1016/0550-3213(87)90691-2

\bibitem{Higuchi:1989gz}
  A.~Higuchi,
  Nucl.\ Phys.\ B {\bf 325}, 745 (1989).
  doi:10.1016/0550-3213(89)90507-5


\bibitem{Crisostomi2012} M. Crisostomi, D. Comelli, and Luigi Pilo, JHEP {\bf 06}, 085 (2012).
\bibitem{Solomon2014} A. Solomon, Y. Akrami, and T. Koivisto, JCAP {\bf1410}, 066 (2014).
\bibitem{Koenning2014} F. K\"{o}nnig, Y. Akrami, L. Amendola, M. Motta, and A. Solomon, Phys. Rev. D {\bf90}, 124014 (2014).
\bibitem{Lagos2014} M. Lagos, and P. Ferreira, JCAP {\bf 12}, 026 (2014).
\bibitem{Comelli2014prd} D. Comelli, M. Crisostomi, and L. Pilo, Phys. Rev. D {\bf 90}, 084003 (2014).
\bibitem{Koenning2014prd2} F. K\"{o}nnig and L. Amendola, Phys. Rev. D {\bf 90}, 044030 (2014).
\bibitem{Berg2012} M. Berg, I. Buchberger, J. Enander, E. Mortsell, and S. Sjors, JCAP {\bf1212}, 021 (2012).

\bibitem{Lagos2016JCAP} M. Lagos, and J. Noller, JCAP {\bf01}, 023 (2016).
\bibitem{Akrami20131} Y. Akrami, T. S. Koivisto, D. F. Mota and M. Sandstad, “Bimetric gravity doubly coupled to matter: theory and cosmological implications”, JCAP {\bf1310}, 046 (2013).
\bibitem{Rham2015} C. de Rham, L. Heisenberg and R. H. Ribeiro, Class. Quant. Grav. {\bf32}, 035022 (2015).
\bibitem{Koenning20141} F. K\"{o}nnig, A. Patil, and L. Amendola, JCAP 03, 029 (2014).
\bibitem{Akrami2014} Y. Akrami, S. F. Hassan, F. K\"{o}nnig, A.Schmidt-May, and A. Solomon, Phys. Lett. B {\bf 748}, 37 (2015).
\bibitem{Felice2014} A. Felice, A. Gumrukcuoglu, S. Mukohyama, N. Tanahashi, and T. Tanaka, JCAP {\bf 1406}, 037 (2014) .
\bibitem{Strauss2012} M. von Strauss, A. Schmidt-May, J. Enander, E. M\"{o}rtsell, and S. Hassan, JCAP {\bf 1203}, 042 (2012).
\bibitem{Volkov2012} M. Volkov, JHEP {\bf 01}, 035 (2012).
\bibitem{Katsuragawa2014} T. Katsuragawa, Phys.Rev. D {\bf 89}, 124007 (2014).



\bibitem{Koenning2015} F. K\"{o}nnig, Phys. Rev. D {\bf 91}, 104019 (2015).

\bibitem{Mortsell2017} E. Mortsell, arXiv:1701.00710 [gr-qc].
\bibitem{Lagos2016} M. Lagos, P. Ferreira, arXiv:1610.00553 [gr-qc].
\bibitem{Apolo2016} L. Apolo, S. Hassan, A. Lundkvist, Phys. Rev. D {\bf 94}, 124055 (2016).
\bibitem{Sakakihara2016} Y. Sakakihara, and T. Tanaka, JCAP {\bf 09}, 033 (2016).
\bibitem{Kimura2016} R. Kimura, T. Tanaka, K. Yamamoto, and Y. Yamashita, Phys. Rev. D {\bf 94}, 064059 (2016).
\bibitem{Do2016} T. Q. Do,  Phys. Rev. D {\bf 94}, 044022 (2016).
\bibitem{Aoki2016} K. Aoki, K. Maeda, and M. Tanabe, Phys. Rev. D {\bf 93}, 064054 (2016).
\bibitem{Gao2016} X. Gao, L. Heisenberg,  JCAP {\bf03}, 043 (2016).
\bibitem{Darabi2016} F. Darabi, and M. Mousavi, Phys. Lett. B, {\bf 761}, 269 (2016).

\bibitem{Akrami2015} Y. Akrami, T. S. Koivisto, and A. R. Solomon, Gen. Rel. Grav. {\bf 47}, 1838 (2015).
\bibitem{Hassan2014} S. Hassan, A. Schmidt-May, and M. von Strauss, Int. J. Mod. Phys. D, {\bf 23}, 1443002 (2014).
\bibitem{Rham20142} C. de Rham, L. Heisenberg, and R. H. Ribeiro, Phys.Rev. D {\bf 90}, 124042 (2014).
\bibitem{Cusin2014} G. Cusin, J. Fumagalli, and M. Maggiore, JHEP {\bf 09}, 181 (2014).
\bibitem{Noller2015} J. Noller and S. Melville, JCAP {\bf 01}, 003 (2015).
\bibitem{Comelli2014} D. Comelli, F. Nesti, and L. Pilo, JCAP {\bf 05}, 036 (2014).
\bibitem{Rham20143} C. de Rham, M. Fasiello, and A. J. Tolley, Int. J. Mod. Phys. D {\bf 23}, 1443006 (2014).

\bibitem{Falck:2014jwa}
  B.~Falck, K.~Koyama, G.~b.~Zhao and B.~Li,
  JCAP {\bf 1407}, 058 (2014)
  doi:10.1088/1475-7516/2014/07/058
  [arXiv:1404.2206 [astro-ph.CO]].

\bibitem{Barreira:2013eea}
  A.~Barreira, B.~Li, W.~A.~Hellwing, C.~M.~Baugh and S.~Pascoli,
  JCAP {\bf 1310}, 027 (2013)
  doi:10.1088/1475-7516/2013/10/027
  [arXiv:1306.3219 [astro-ph.CO]].

\bibitem{Li:2013tda}
  B.~Li, A.~Barreira, C.~M.~Baugh, W.~A.~Hellwing, K.~Koyama, S.~Pascoli and G.~B.~Zhao,
  JCAP {\bf 1311}, 012 (2013)
  doi:10.1088/1475-7516/2013/11/012
  [arXiv:1308.3491 [astro-ph.CO]].

\bibitem{Barreira:2013xea}
  A.~Barreira, B.~Li, C.~M.~Baugh and S.~Pascoli,
  JCAP {\bf 1311}, 056 (2013)
  doi:10.1088/1475-7516/2013/11/056
  [arXiv:1308.3699 [astro-ph.CO]].


\bibitem{Ade:2015rim}
  P.~A.~R.~Ade {\it et al.} [Planck Collaboration],
  Astron.\ Astrophys.\  {\bf 594}, A14 (2016)
  doi:10.1051/0004-6361/201525814
  [arXiv:1502.01590 [astro-ph.CO]].

\bibitem{Comelli2012} D. Comelli, M. Crisostomi, F. Nesti, and L. Pilo, JHEP {\bf 1203},  067 (2012) [arXiv:1111.1983].
\bibitem{Tamanini2014} N. Tamanini, E. Saridakis, and T. Koivisto, JCAP {\bf 1402}, 015 (2014).


\bibitem{Khosravi:2012rk}
  N. Khosravi, H. R. Sepangi and S. Shahidi,
  Phys. Rev. D {\bf 86}, 043517 (2012)
  doi:10.1103/PhysRevD.86.043517
  [arXiv:1202.2767 [gr-qc]].

\bibitem{Konnig:2014xva}
  F. K\"{o}nnig, Y. Akrami, L. Amendola, M. Motta and A. R. Solomon,
  Phys. Rev. D {\bf 90}, 124014 (2014)
  doi:10.1103/PhysRevD.90.124014
  [arXiv:1407.4331 [astro-ph.CO]].

\bibitem{Ma:1995ey}
  C.~P.~Ma and E.~Bertschinger,
  Astrophys. J.  {\bf 455}, 7 (1995)
  doi:10.1086/176550
  [astro-ph/9506072].

\bibitem{Abazajian:2008wr}
  K.~N.~Abazajian {\it et al.} [SDSS Collaboration],
  Astrophys.\ J.\ Suppl.\  {\bf 182}, 543 (2009)
  doi:10.1088/0067-0049/182/2/543
  [arXiv:0812.0649 [astro-ph]].





























\end{thebibliography}
\end{document}